\documentclass[12pt]{iopart}

\usepackage[numbers,sort&compress]{natbib}
\usepackage{iopams}

\expandafter\let\csname equation*\endcsname\relax

\expandafter\let\csname endequation*\endcsname\relax
\usepackage[cmex10]{amsmath}

\usepackage{algorithmic}
\usepackage{array}
\usepackage[caption=false,font=footnotesize]{subfig}
\usepackage{fixltx2e}
\usepackage{stfloats}
\usepackage{graphicx}
\usepackage{multirow}
\usepackage{threeparttable}

\begin{document}
\title[ERASE for removal of EMG artifacts]{Electromyogram (EMG) Removal by Adding Sources of EMG (ERASE) - A novel ICA-based algorithm for removing myoelectric artifacts from EEG - Part 2}

\author{Yongcheng Li$^1$, Po T. Wang$^2$,
Mukta P. Vaidya$^{3,4,5}$, Charles Y. Liu $^{6,7,8}$, Marc W. Slutzky $^{3,4,5}$ and An H. Do$^1$}

\address{$^1$ Department of Neurology, University of California, Irvine, CA 92697 USA}
\address{$^2$ Department of Biomedical Engineering, University of California, Irvine, CA 92697, USA}
\address{$^3$ Department of Neurology, Northwestern University, Chicago, Illinois, USA}
\address{$^4$ Department of Physiology, Northwestern University, Chicago, Illinois, USA}
\address{$^5$ Department of Physical Medicine and Rehabilitation, Northwestern University, Chicago, Illinois, USA}
\address{$^6$ Department of Neurosurgery, University of Southern California, CA, USA}
\address{$^7$ Rancho Los Amigos National Rehabilitation Center, CA, USA}
\address{$^8$ Neurorestoration Center, University of Southern California, CA, USA}

\ead{and@uci.edu; yongchel@uci.edu}

\begin{abstract}
Recent studies have shown the ability to record high-$\gamma$ signals (80 - 160 Hz) in electroencephalogram (EEG) from traumatic brain injury (TBI) patients who have had hemicraniectomies. However, extraction of the movement-related high-$\gamma$ remains challenging due to a confounding bandwidth overlap with surface electromyogram (EMG) artifacts related to facial and head movements. In part 1, we described an augmented independent component analysis (ICA) approach for removal of EMG artifacts from EEG, and referred to as \emph{EMG Reduction by Adding Sources of EMG (ERASE)}. Here, we tested this algorithm on  EEG recorded from six TBI patients with hemicraniectomies while they performed a thumb flexion task. ERASE removed a mean of 52 $\pm$ 12\% (mean $\pm$ S.E.M) (maximum 73\%) of EMG artifacts. In contrast, conventional ICA removed a mean of 27 $\pm$ 19\% (mean $\pm$ S.E.M) of EMG artifacts from EEG. In particular, high-$\gamma$ synchronization was significantly improved in the contralateral hand motor cortex area within the hemicraniectomy site after ERASE was applied. 
A more sophisticated measure of high-$\gamma$ complexity is the fractal dimension (FD). Here, we computed the FD of EEG high-$\gamma$ on each channel. Relative FD of high-$\gamma$ was defined as that the FD in move state was subtracted by FD in idle state. We found relative FD of high-$\gamma$ over hemicraniectomy after applying ERASE were strongly correlated to the amplitude of finger flexion force. Results showed that significant correlation coefficients across the electrodes related to thumb flexion averaged $\sim$0.76, while the coefficients across the homologous electrodes in non-hemicraniectomy areas were nearly 0. After conventional ICA, a correlation between relative FD of high-$\gamma$ and force remained high in both hemicraniectomy areas (up to 0.86) and non-hemicraniectomy areas (up to 0.81). Across all subjects, an average of 83\% of electrodes significantly correlated with force was located in the hemicraniectomy areas after applying ERASE. After conventional ICA, only 19\% of electrodes with significant correlations were located in the hemicraniectomy. These results indicated that the new approach isolated electrophysiological features during finger motor activation while selectively removing confounding EMG artifacts. This approach removed EMG artifacts that can contaminate high-$\gamma$ activity recorded over the hemicraniectomy.
\end{abstract}
\noindent{EMG artifact; EEG; artifacts removal; ICA; high-$\gamma$; TBI; BMI/BCI}
\maketitle
\normalsize

\section{Introduction}

Conventional EEG has a poor signal-to-noise ratio in the high-$\gamma$ band due to the low-pass filter characteristics of the skull and scalp \citep{47,48}. Meanwhile, movement and force are strongly encoded in the high-$\gamma$ band activity from brain signals \citep{1,2,68,69,70,71,72,73,74}. Traumatic brain injury (TBI) patients with hemicraniectomy may be a useful model for human electrophysiology with high bandwidth and spatiotemporal resolution \citep{5,62}. In particular, substantial high-$\gamma$ band power can be detected in these patients' electroencephalogram (EEG) due to the absence of the skull in the hemicraniectomy area (referred to as hEEG) \citep{50,51}. 
However, the extraction of the high-$\gamma$ band features from hEEG in TBI patients remains challenging due to the large confounding spectral overlap between the high-$\gamma$ band of EEG and EMG, which is primarily caused by facial and head movement and has broad bandwidth\citep{9,49,54,55}. 

To surmount this challenge, we tested the ability of our novel approach, referred to as \emph{EMG Reduction by Adding Sources of EMG (ERASE)} (detailed in Part 1), to minimize the EMG artifacts in hEEG. We applied the technique to EEG from 6 TBI patients while they performed isometric finger flexion, and found the residual high-$\gamma$ signal in hEEG was correlated to force level.

\section{Methods}
\subsection{Experiments}
This study was approved by the Institutional Review Boards of the University of California, Irvine, Northwestern University and the Rancho Los Amigos National Rehabilitation Center (RLANRC). TBI patients with hemicraniectomy and mild to moderate weakness on the contralesional hand were recruited for our study. Subjects were fitted with a 128-electrode EEG cap (ActiCap, Brain Products, Gilching, Germany) and asked to perform a thumb flexion task on the contralesional side while the EEG signals were acquired at 2000 Hz (Neuroport, Blackrock Microsystems, Salt Lake City, UT). The thumb flexion force was simultaneously measured by a load cell sensor. The subjects were instructed to flex their thumbs on the affected side to apply varying levels of force on the load cell sensor to move a computer cursor to acquire targets in a 1D, random-force target task. Each flexion event was set to be 2 seconds long (target displayed for 2 seconds) with a 3-5 seconds interval between each event. Subjects nominally performed 20 
thumb flexions in each 120-s  run at RLANRC and 53 
thumb flexions in each 300s-long run at Northwestern University. All of the trails were used in the analysis described below.

\subsection{Experimental data processing}

Only simulated EMG was used as reference EMG artifacts, as described in Section 2.2.1 in Part 1 in Supplementary Materials. The EEG from TBI patients with hemicraniectomy were subjected to the same processing approach described in Section 2.3.2 in Part 1 in Supplementary Materials. Briefly, the EEG and EMG data were combined (simulated EMG acted as separate electrodes and were not mixed with any EEG signal), and the combined EEG/EMG data were subjected to a 3-200 Hz 3$^{\text{rd}}$-order bandpass filter. All trials were identified and extracted from the EEG/EMG data. 
Each trial was defined as 1-s idle time (remaining 2-4 s of idle time discarded) followed by 2-s movement. The extracted trials were concatenated. Since EEG always includes other unexplained noise, leading to long-term EEG non-stationary, ICA decomposition was applied to concatenated EEG trials (FastICA algorithm, EEGLAB toolbox \citep{15}) which contained the entire broad bandwidth. The automated artifacts IC rejection process described in Section 2.1.2 in Part 1 was applied. Short-time Fourier transform was applied to EEG trials and the signal power in different frequency bands ($\mu$ band: 8 to 12 Hz, high-$\gamma$ band: 80 to 160 Hz) were compared across all conditions (baseline, ERASE with simulated EMG, conventional ICA). All the data were z-scored after the time-frequency decomposition. Also, we calculated the percent reduction for high-$\gamma$ band across all electrodes in non-hemicraniectomy areas by using equation (11) in Part 1.

\subsection{Signal-to-noise ratio (SNR) calculation}
To evaluate if the electrophysiological features and information content underlying movement were retained after ERASE, the signal-to-noise ratio (SNR) between movement and idle was calculated for electrodes expected to be related to thumb movements, including \emph{C3, C5, C1, FCC5h, FCC3h, CCP5h, and CCP3h} for subjects with left-sided hemicraniectomy, or \emph{C4, C2, C6, FCC6h, FCC4h, CCP4h, and CCP6h} for those with right-sided hemicraniectomy. These electrodes will be subsequently referred to as \textit{hand motor electrodes} (see Supplementary Fig. 11). Note that the homologous electrodes on the non-hemicraniectomy side will be referred to as the contralesional electrodes. Generally, most of hand motor electrodes were located in the hemicraniectomy areas. Here, we treated the high-$\gamma$ power during movement as signal and the baseline high-$\gamma$ power during idle as noise. The SNR for each trial before and after applying ICA was calculated as in \citep{12}. The segmented trials were concatenated as described in Section 2.3.2 in Part 1 in Supplementary and the resulting time series was referred to as $X(t)$. For each electrode, the EEG was subjected to bandpass filtered from 80-160 Hz (4th order Butterworth filters) to extract the high-$\gamma$ band, referred to as $X_{\gamma}(t)$. The resulting signal was then squared to obtain the instantaneous high-$\gamma$ band power, $X_{\gamma}^2(t)$. $X_{\gamma}^2(t)$ was low-pass filtered (4 Hz, 4th order Butterworth filter) to create the envelope signal, $P_\gamma$. The average power of high-$\gamma$ band during movement and idling time were calculated, denoted as $\overline{P_{m,\gamma}}$ and $\overline{P_{i,\gamma}}$, respectively. The SNR was defined as $10\times\log_{10}{(\overline{P_{m,\gamma}}/\overline{P_{i,\gamma}})}$. This processing was also applied to $\mu$ band. We calculated the SNR at both of $\mu$ and high-$\gamma$ bands for all the available trials of each subject.
A Wilcoxon rank-sum test was used to compare the SNR between all pair combinations of conditions (baseline, ERASE with simulated EMG, conventional ICA) for both bands.

\subsection{EEG high-$\gamma$ verification}
To measure the success of ERASE, we assessed the extent to which high-$\gamma$ movement-related information content was retained after it has been applied. Previous studies have demonstrated that high-$\gamma$ band power increased during muscle activation or movement \citep{2,3,12,64,65,66} and that fractal dimension (FD) can be used as a measure of brain activity by quantifying the complexity of EEG \citep{37,38}. Also, the FD of EEG signal has been shown to change with the level of force \citep{34}.  Since relative FD (change in FD between idle and move states) can be an indicator of cortical activity modulation during movement \citep{34}, the correlation between thumb flexion force and the high-$\gamma$ relative FD will be used to assess if the recovered high-$\gamma$ in hEEG still retains encoding of movement. The correlation between the high-$\gamma$ relative FD and the thumb flexion force was calculated as follows.
First, the FD of the EEG high-$\gamma$ was calculated as follows for the move and idle epochs in each trial, respectively \citep{34,35,36}:
\begin{equation}\label{6}
	FD=\frac{ln(N-1)}{ln(N-1)+ln(d/L)}
\end{equation}
where N is the total number of time points to be analyzed for each epoch (2000 for idle epoch, 4000 for movement epoch), L is the sum of the Euclidean distances between successive data vectors, and d is the Euclidean distance between the first data vector and the vector that provides the farthest distance. Data vector was composed of successive time points, which were less than $\frac{N}{2}$. Any adjacent data vectors included no overlapped time points. Since the FD value was dependent on quantization units in this algorithm \citep{35}, 1 ms was chosen as the quantization unit of time (i.e., one data vector was composed of two time points, hence, multiple data vectors were included in each epoch), and 1$\mu$V as that of the EEG potential. Here, we calculated the Euclidean distances for all the data vectors in the linear space.
Next, the relative FD, defined as that FD value during idle epoch subtracted from that during move epoch, was calculated for each trial.

 For each subject, the mean force during the move epoch for each trial was calculated. Then the maximum and minimum of these mean force were found. Since the resolution of high-$\gamma$ in hEEG was not sufficient to precisely decode continuous force, the mean force was evenly discretized into 10 force levels from minimum to maximum. Subsequently, the relative FD values were averaged over the trials at each force level for each subject. The correlation coefficient (R, $|R|$ denoted the absolute value of correlation coefficient) between force level and relative FD values for all electrodes was calculated with Pearson Correlation. The significance of correlation was calculated by considering the correlation coefficients as a t distribution. The t-value can be calculated as below \citep{67}:
\begin{equation}\label{7}
t=\frac{|R|*\sqrt{N-1}}{\sqrt{1-R^2}}
\end{equation}
where R is correlation coefficient, N is the sample number.
In our work, the correlation coefficients with no significance (P $>$ 0.05) were set as zero, and the ones with significance denoted as significant R-value (or significant $|R|$-value for absolute value).

\begin{table}[!b]
\centering
\caption{Subject Demographics}\label{table1}
\begin{threeparttable}
\begin{tabular}{p{0.06\columnwidth}p{0.08\columnwidth}p{0.08\columnwidth}p{0.2\columnwidth}p{0.3\columnwidth}}
  \hline
 Subject & Age & Sex & HA side & Contralesional hand\\
    S1 & 23 & Female & Left & Right hand \\
    S2 & 34 & Male & Left & Right hand \\
    S3 & 30 & Male & Right & Left hand \\
    S4 & 40 & Male & Right & Left hand \\
    S5 & 29 & Male & Right & Left hand \\
    S6 & 56 & Female & Left & Right hand \\
    \hline
\end{tabular}

   \begin{tablenotes}
        \small
        \item HA was the hemicraniectomy areas.
       \end{tablenotes}
    \end{threeparttable}
\end{table}

\section{RESULTS}
A total of 58 sessions (including 1771 trials) were collected from 6 TBI patients with hemicraniectomy (subject demographics summarized in Table 1).

\begin{figure}[!t]
  \centering
  \vspace{-0.1in}
  \includegraphics[width=\textwidth]{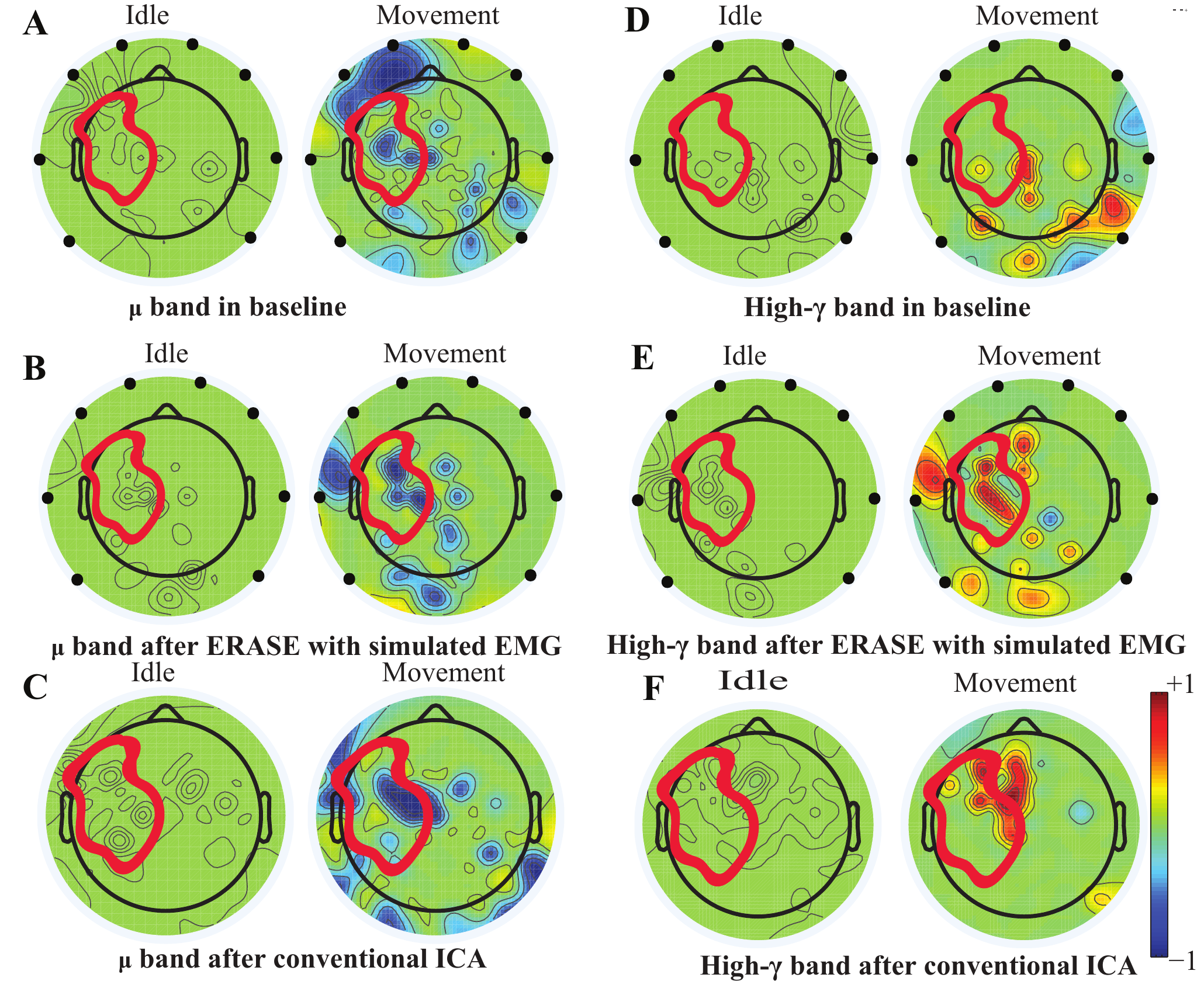}
  \caption{Brain topography maps of z-scored $\mu$ and high-$\gamma$ power before and after artifacts rejection with ERASE and conventional ICA on the Subject 1. 
  	Only electrodes whose z-scored power of $\mu$ and high-$\gamma$ during idle time and movement were significantly different were shown (Wilcoxon rank-sum test, details were described in Section 2.3.2 in part 1). P-value for significant difference was 0.05. The black dots outlined the position of the added virtual electrodes. The red outline in each subfigure denoted the HA. }\label{fig.1}
\vspace{-0.1in}
\end{figure}

\begin{figure}[!t]
  \centering
   \vspace{-0.1in}
  \includegraphics[width=\textwidth]{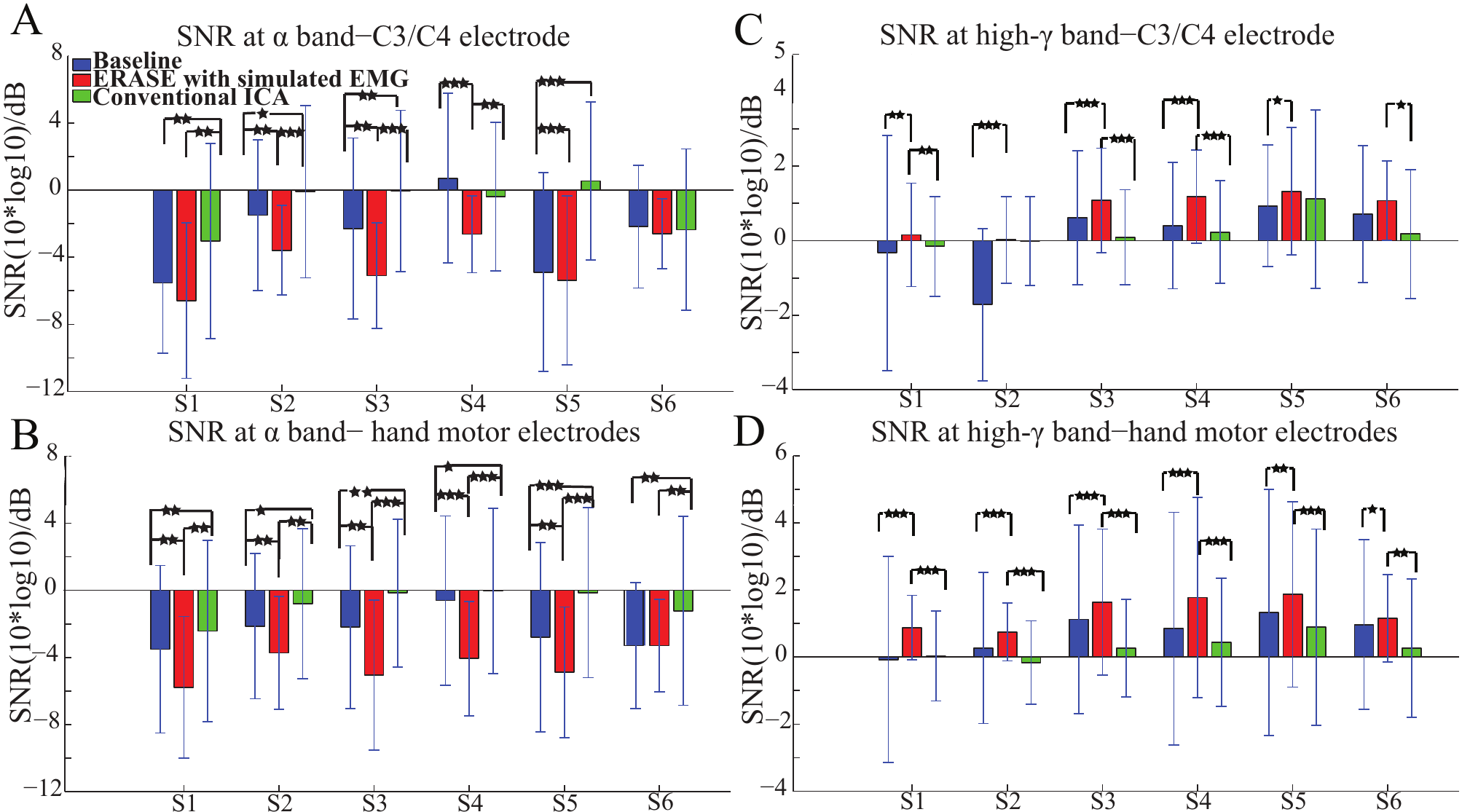}
  \caption{Mean SNR ($\pm$ standard deviation (S.D)) across all available trials for each subject in different conditions (baseline, after ERASE and after conventional ICA). S1 was the abbreviation for Subject 1, and so on. The asterisks indicated the significant differences between the two datasets, and the significance level=***p$<$0.001, level=**p$<$0.01, and level=*p$<$0.05 (Wilcoxon rank-sum test).  \textbf{A.} SNR of $\mu$ band. Data were from the C3/C4 electrode. \textbf{B.} SNR of $\mu$ band. Data were from hand motor electrodes. \textbf{C.} SNR of high-$\gamma$ band. Data were from the C3/C4 electrode. \textbf{D.} SNR of high-$\gamma$ band. Data were from hand motor electrodes.}\label{fig.2}
  \vspace{-0.1in}
\end{figure}

\subsection{Brain features evaluation}

 After processing of EEG as described in Section 2.3.2 in part 1, representative examples showed that the z-scored EEG high-$\gamma$ power in the non-hemicraniectomy areas (NHAs) was reduced while $\mu$ desynchronization (negative SNR value) and high-$\gamma$ synchronization (increased SNR) in the motor-related areas (in the hemicraniectomy areas (HAs)) were apparent after ERASE (Fig. 1 and Supplementary Figs. 1 to 5). In baseline, the average Z-scored high-$\gamma$ power was not different between HA and non-HA across all subjects. Only after ERASE did the average Z-scored high-$\gamma$ power during movement across all the electrodes in the HAs became significantly larger than that across all the electrodes in the NHAs (P-value $<$ 0.05, Wilcoxon rank-sum test, Table 2). The z-scored high-$\gamma$ power during movement in the NHA was reduced by an average of 52.03 $\pm$ 12.08\% (mean $\pm$ S.E.M) across all subjects (Table 2), indicating that artifacts were removed. 
 On the other hand, the corresponding reduction in non-HA for conventional ICA condition averaged 26.50 $\pm$ 18.91\% (mean $\pm$ S.E.M) across all subjects (Table 2).

The high-$\gamma$ synchronization described above was seen in the C3 (for subjects with left hemicraniectomy, right thumb flexion) /C4 (subjects with right hemicraniectomy, left thumb flexion) electrode in 4 out of 6 subjects in baseline. When considering all hand motor electrodes, high-$\gamma$ synchronization was observed in 5 out of 6 subjects in baseline (Fig. 2 C and D).
 The high-$\gamma$ synchronization at the C3/C4 electrode was significantly larger after ERASE in 5 out of 6 subjects compared to baseline condition. However, the C3/C4 high-$\gamma$ synchronization after conventional ICA was never significantly larger than those in baseline (Fig. 2C). The high-$\gamma$ synchronization from hand motor electrodes was significantly larger after ERASE compared to those both in baseline and after conventional ICA for all the subjects. However, except in Subject 1, the high-$\gamma$ synchronization from hand motor electrodes in the conventional ICA condition was smaller than those in baseline (Fig. 2D). These findings suggested that high-$\gamma$ remaining in the HAs after ERASE was movement-related EEG, and ERASE was more effective at preserving the EEG features.

In order to further verify these high-$\gamma$ activities were neurogenic and related to movement, the $\mu$ desynchronization was calculated as described in Section 2.3. The $\mu$ desynchronization was present in 5 out of 6 subjects in baseline at C3/C4. After applying ERASE, it was present in all subjects at C3/C4, and the magnitude of desynchronization increased significantly in 4 subjects and was not significantly different in the remaining 2 (Fig. 2 A). In contrast, after conventional ICA, the $\mu$ desynchronization was no longer present in 1 out of 6 subjects. Furthermore, the magnitude of desynchronization was significantly reduced in 4 subjects (Fig. 2 A).
 Across all of the hand motor electrodes, $\mu$ desynchronization was present in all subjects in baseline. After applying ERASE, the magnitude of desynchronization was significantly larger in 5 out of 6 subjects compared to baseline condition (Fig. 2 B). After conventional ICA, the magnitude of desynchronization was significantly reduced in all subjects. The magnitude of desynchronization was always significantly larger after applying ERASE compared to that of the conventional ICA (Fig. 2 B).

\begin{table}[!t]
\centering
\caption{z-scored high-$\gamma$ power in different conditions for each subject.}\label{table2}
\begin{threeparttable}
\begin{tabular}{p{0.31\columnwidth}p{0.06\columnwidth}p{0.06\columnwidth}p{0.06\columnwidth}p{0.06\columnwidth}p{0.06\columnwidth}p{0.06\columnwidth}}
  \hline
 Subject &S1 & S2 & S3 & S4 & S5 & S6\\
    Number of trials & 261 & 97 & 864 & 230 & 209 & 110\\
    \hline
    & \multicolumn{6}{c}{\textbf{Baseline}}\\
    Mean z-scored in NHA& 0.26 & 0.13 & 0.21 & 0.24 & 0.26 & 0.15\\
    Mean z-scored in HA& 0.25 & 0.12 & 0.21 & 0.23 & 0.35 & 0.13\\
    P-value& 0.35 & 0.23 & 0.47 & 0.42 & 0.67 & 0.62\\
    \hline
    & \multicolumn{6}{c}{\textbf{ERASE with simulated EMG}}\\
    Mean z-scored in NHA& 0.16 & 0.06 & 0.12 & 0.11 & 0.12 & 0.04\\
    Mean z-scored in HA& 0.24 & 0.10 & 0.19 & 0.19 & 0.34 & 0.12 \\
      P-value& 0.02 & 0.03 & 0.04 & 0.04 & 0.03 & 0.03\\
  \hline
  & \multicolumn{6}{c}{\textbf{Conventional ICA}}\\
    Mean z-scored in NHA& 0.10 & 0.11 & 0.16 & 0.18 & 0.24 & 0.11\\
    Mean z-scored in HA& 0.14 & 0.12 & 0.16 & 0.15 & 0.2 & 0.12\\
   P-value& 0.12 & 0.63 & 0.57 & 0.17 & 0.96 & 0.61\\
  \hline
\end{tabular}
\begin{tablenotes}
        \small
        \item Mean z-score denoted the average z-scored high-$\gamma$ power during movement over all the available trials. The mean z-scored high-$\gamma$ power in the HAs got averaged across all the electrodes in the HAs. Meantime, The mean z-scored high-$\gamma$ power in the NHAs got averaged across all the electrodes in the NHAs
        \item HA: Hemicraniectomy area. NHA: Non-HA
        \item P-value: comparison of z-scored high-$\gamma$ power between HAs and NHAs for each subject (Wilcoxon rank-sum test).
      \end{tablenotes}
    \end{threeparttable}
    \vspace{-0.1in}
\end{table}

\subsection{EEG High-$\gamma$ verification}

A strong correlation between thumb flexion force and the relative FD of high-$\gamma$ was primarily found in the HAs for all the subjects after ERASE (Fig. 3). After conventional ICA, either there were no strong correlations in the HAs (Fig. 3 S1 and S4), or a strong correlation was seen in both HAs and non-HAs (Fig. 3 S2, S3, S5, and S6).

The correlations between relative FD and thumb flexion force for each condition are summarized in Table 3.
In baseline, 27.34\% of electrodes with significant correlation were located within HAs across all subjects.
Subsequently, 83.33\% and 18.99\% of electrodes with significant correlation were located within HAs after applying ERASE and the conventional ICA, respectively (Table 3, Fig. 4 and Supplementary Figs. 6-10). 

In baseline, the average significant $|R|$-value in hand motor electrodes was 0.436 across all subjects (subjects with no significant correlation were treated as though the significant $|R|$-value was 0). This was lower than the significant $|R|$-values on the contralesional electrodes area (0.654 from Table 3). The average significant $|R|$-values within the hand motor electrodes were 0.762 and 0.528 across all subjects for ERASE and conventional ICA conditions, respectively (Table 3).
By comparison, ERASE removed the presence of any correlation in the contralesional electrodes areas except in 1 subject (only one electrode was still correlated, average significant $|R|$ across all the subjects was 0.11). After conventional ICA, the average significant $|R|$-value within contralesional electrodes across all the subjects was 0.26 (Table 3).

\begin{table}[!b]
\centering
\caption{Average of significant $|R|$-values in hand motor electrodes and contralesional electrodes in different conditions.}\label{table3}
\begin{threeparttable}
\begin{tabular}{p{0.38\columnwidth}p{0.04\columnwidth}p{0.04\columnwidth}p{0.04\columnwidth}p{0.04\columnwidth}p{0.04\columnwidth}p{0.04\columnwidth}}
  \hline
 Subject & S1 & S2 & S3 & S4 & S5 & S6\\
 \multicolumn{7}{c}{\textbf{Baseline}}\\
$|R|$-value in hand motor&0.61 &0.76 &0 &0.72 &0.53 &0 \\
SCE number in hand motor&4  &1 &0 &1 &3 &0\\
$|R|$-value in contralesional &0.73 &0.81 &0.77 &0 &0.68 &0.93 \\
SCE number in contralesional&1 &1 &1 &0 &1 &1\\
Total number of SCE &31 & 8 & 14 & 4 & 8 & 21 \\
Proportion of SCE in HA (\%)&35.48 & 37.50 & 14.29 & 25 & 37.5 & 14.29 \\
 \hline
 \multicolumn{7}{c}{\textbf{ERASE with simulated EMG}}\\
$|R|$-value in hand motor &0.62  &0.83 &0.80 &0.74 &0.69 &0.90 \\
SCE number in hand motor&3  &2 &2 &1 &1 &1\\
$|R|$-value in contralesional &0.68 &0 &0 &0 &0 &0 \\
SCE number in contralesional&1 &0 &0 &0 &0 &0 \\
Total number of SCE &12& 5 & 3 & 2 & 4 & 2 \\
Proportion of SCE in HA (\%)&83.33& 100 & 66.67 & 100 & 50 & 100 \\
  \hline
  \multicolumn{7}{c}{\textbf{Conventional ICA}}\\
$|R|$-value in hand motor &0 &0.76 &0.75 &0 &0.80 &0.86 \\
SCE number in hand motor&0  &1 &2 &0 &1 &1\\
$|R|$-value in contralesional &0.73 &0 &0 &0.81 &0 &0 \\
SCE number in contralesional &1 &0 &0 &1 &0 &0 \\
Total number of SCE &10& 10 & 8 & 14 & 11 & 13 \\
Proportion of SCE in HA (\%)&0& 30 & 50 & 0 & 18.19 & 15.38 \\
      \hline
\end{tabular}
    \end{threeparttable}
    \begin{tablenotes}
        \small
        \item[1] SCE was significant correlation electrodes
        \item[2] hand motor denoted hand motor electrodes, which included C3, C5, C1, FCC5h, FCC3h, CCP5h, and CCP3h for right hand, or C4, C2, C6, FCC6h, FCC4h, CCP4h, and CCP6h for left hand. Contralesional denoted the contralesional electrodes, which were the homologous motor electrodes in the non-hemicraniectomy area side.
       \end{tablenotes}
    \vspace{-0.1in}
\end{table}

\begin{figure}[!ht]
  \centering
  \vspace{-0.1in}
  \includegraphics[width=4 in]{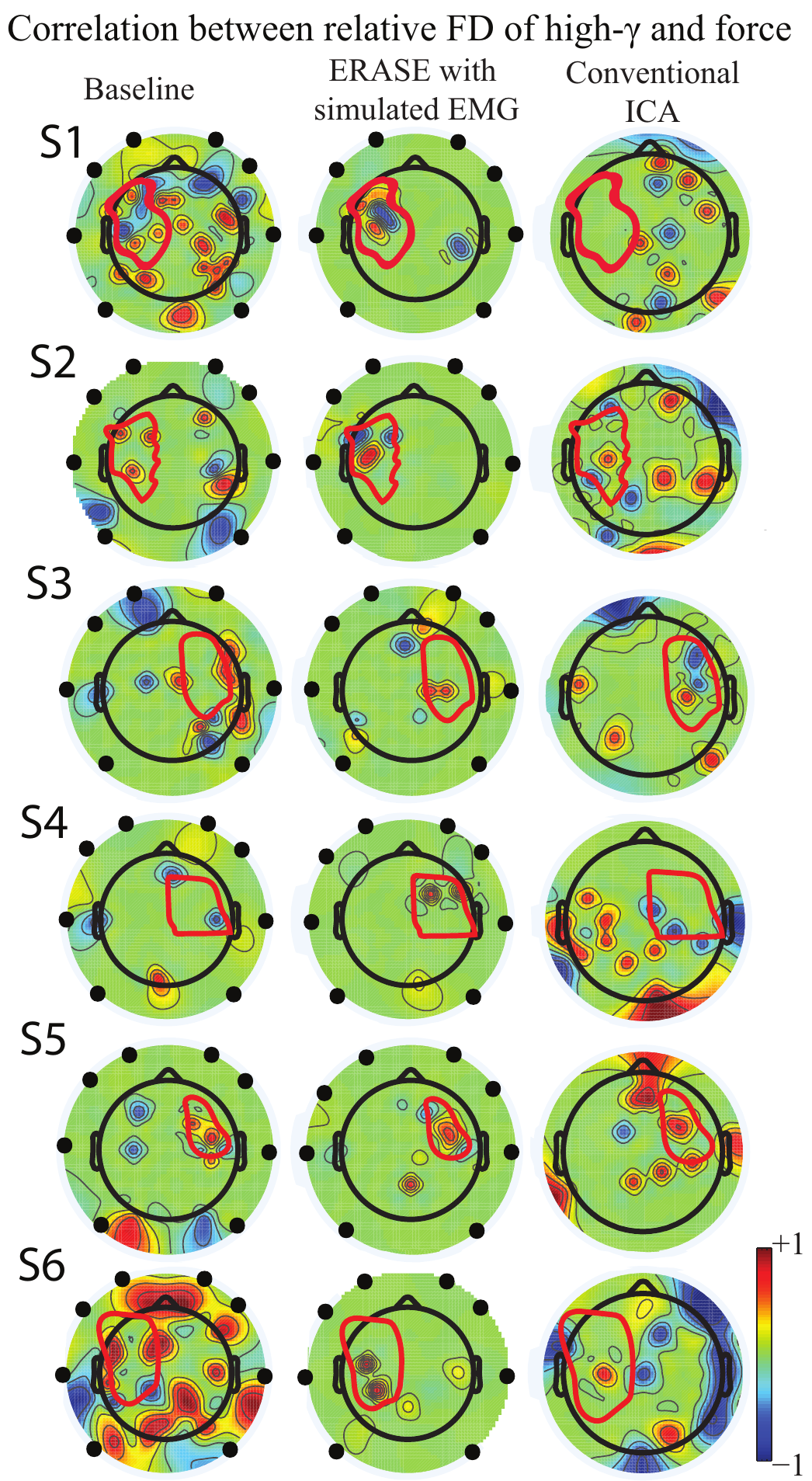}
  \caption{Correlation between the relative FD of high-$\gamma$ and the amplitude of thumb flexion force in different conditions (baseline, after ERASE and after conventional ICA). Only the significant correlation coefficients were showed, and other values were set to 0. The black dots outlined the position of the added virtual electrodes. The red outline denoted the hemicraniectomy areas. \textbf{S1-S6} denoted Subject 1-6, respectively.}\label{fig.3}
\vspace{-0.1in}
\end{figure}

\section{Discussion}
\begin{figure}[!hb]
  \centering
  \vspace{-0.1in}
  \includegraphics[width=\textwidth]{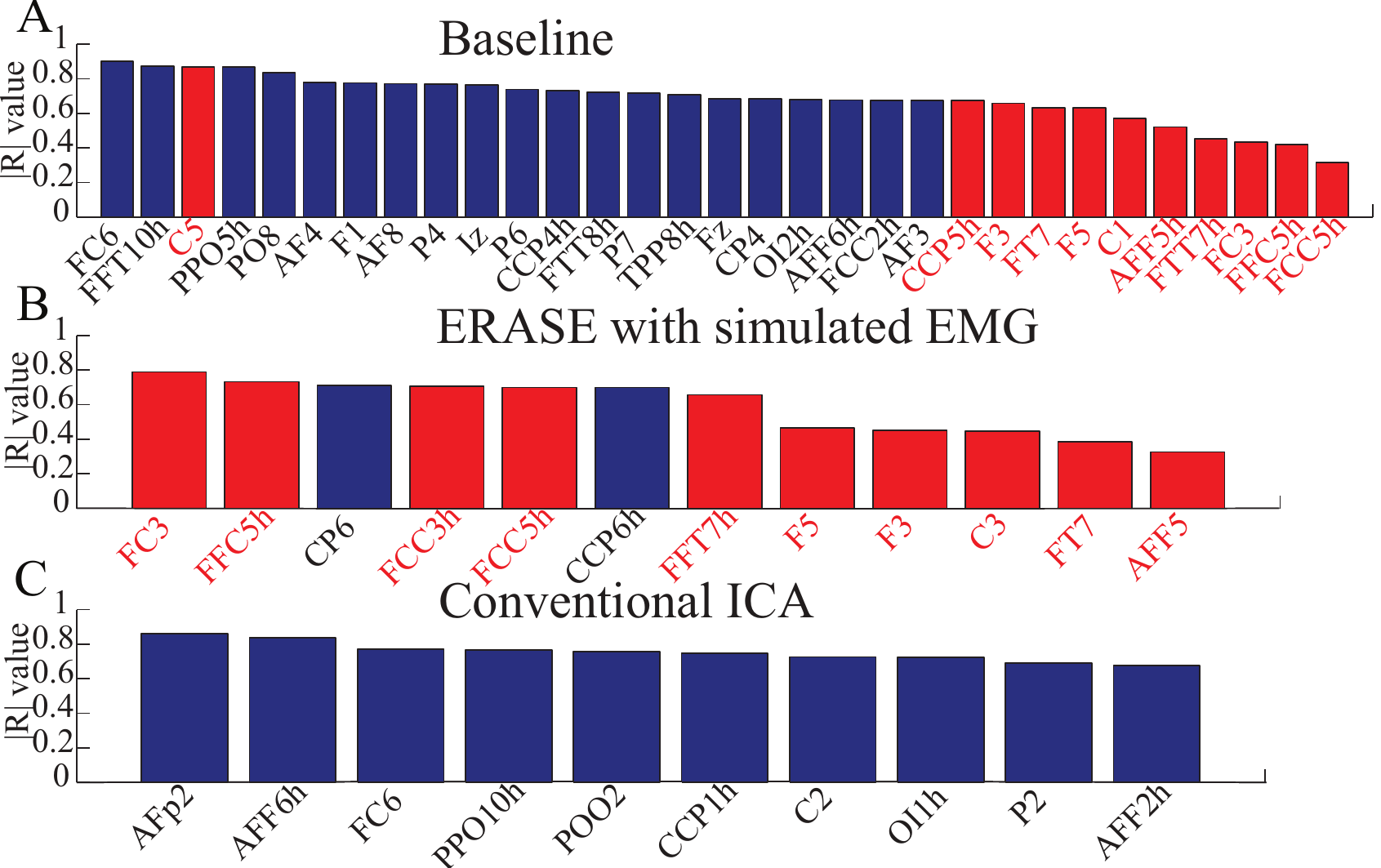}
  \caption{Bar graphs for showing the electrodes with significant correlation in three conditions (baseline, after ERASE with simulated EMG and after running conventional ICA). Here, the correlation coefficients were absolute values from 0 to 1. Data were from Subject 1. Blue bars denoted the electrodes with significant correlation in the NHAs, and red bars were the ones in HAs. \textbf{A.} the electrodes with significant correlation in baseline. \textbf{B.} the electrodes with significant correlation after ERASE with simulated EMG. \textbf{C.} the electrodes with significant correlation after conventional ICA. }\label{fig.4}
\vspace{-0.1in}
\end{figure}
Our novel approach, ERASE, was demonstrated to be an effective tool at removing EMG artifacts from EEG during thumb flexion by TBI patients with HAs while preserving the expected underlying EEG features. Specifically, both $\mu$ desynchronization and high-$\gamma$ synchronization during movement in TBI patients were found in anatomically expected areas, while $\mu$ desynchronization was not seen on the contralesional side. This indicated that the high-$\gamma$ synchronization in hEEG can be more confidently interpreted as cortical activity and not EMG (Fig. 1, Supplementary Figs. 1-5). Also, the SNR of high-$\gamma$ and $\mu$ were both significantly improved compared to conventional ICA and baseline conditions (Fig. 2). The relative FD of high-$\gamma$ was strongly correlated to the thumb flexion force primarily within the HAs after applying our new approach (Figs. 3, 4 and Supplementary Figs. 6-10). These results indicated that ERASE can help to isolate cortical sources of high-$\gamma$ by significantly reducing the presence of EMG artifacts.


Information about the primary kinetics appeared to be preserved in the high-$\gamma$ signal after ERASE. At baseline, high correlation between high-$\gamma$ relative FD and thumb flexion force was similarly present in both HA and NHAs. After applying ERASE, hand motor electrodes located in the HAs typically had the strongest positive correlation across all the subjects, indicating that the new algorithm selectively removed EMG artifacts from the NHAs. This implied that there were fewer artifacts in the HA after ERASE.
Given that ERASE helps to isolate electrophysiological features during hand motor movements while removing confounding EMG artifacts, the justification for using TBI patients with hemicraniectomy as a model for studying high-bandwidth EEG signals is bolstered.

Our recent study on EEG from TBI patients with hemicraniectomy demonstrated that the high-$\gamma$ over HAs was capable of effectively decoding the thumb flexion force \citep{62}. In this study, we demonstrated that high-$\gamma$ over HAs may contain information about thumb flexion force, as demonstrated by a strong correlation between relative FD of high-$\gamma$ and flexion force. Anecdotally, neither this study nor our prior work showed a strong direct correlation between hEEG high-$\gamma$ power and thumb flexion force as seen in ECoG signals \citep{40}. Most likely, the presence of scalp still interferes with the high-$\gamma$ signal enough that such robust features are not preserved when compared to subdurally \citep{3,4} and epidurally \citep{68} recorded ECoG signals. Alternatively, the removal of EMG may also be overly aggressive. More specifically, while EMG is effectively removed, a portion of neurogenic high-$\gamma$ component was also simultaneously removed as well. This highlights a potential tradeoff between EMG removal and neural information preservation in ERASE. In the future, this tradeoff may potentially be minimized by selecting the rejection threshold in a manner that jointly optimizes high-$\gamma$ encoding as well as EMG reduction (similar to Section 2.1.2 in part 1). In another study on TBI patients with hemicraniectomy \citep{5}, both $\mu$ desynchronization and high-$\gamma$ synchronization underlying movement can be found in hEEG. In our work, we not only found that this modulation was associated with movements (Fig. 1, Supplementary Figs. 1-5, Fig. 2 and Supplementary Figs. 6-11), but also showed that there was a strong correlation between the relative FD of high-$\gamma$ and the amplitude of thumb flexion force after employing ERASE. 

Even after ERASE with simulated EMG, the overall reduction of high-$\gamma$ power was 52\% (Table 2), and there was still one subject with EEG electrode in the NHA with a strong correlation between the high-$\gamma$ relative FD and thumb flexion force (Table 3). These findings indicated that not all of the EMG artifacts were easily removed. This is likely due to the fact that simulated EMG artifacts cannot precisely mimic real EMG artifacts since many components of EMG are still difficult to simulate (e.g. the neural sources of real EMG artifacts). We hypothesize that using real EMG as the reference artifacts may lead to better EMG artifact rejection. However, since real EMG is not always available or possible to collect, future work will involve developing an algorithm that further improves upon the EMG removal. This may involve improving the learning algorithms, such as adaptive system recognition and system recognition with artificial neural network, to characterize the EMG. In addition, future work will involve making the technique more computationally efficient, such that it can be used in real time for neurorehabilitation applications, such as in BCI systems.

\section{Conclusion}
Our new approach, ERASE, as described in part 1 can also be applied to effectively remove EMG artifacts from hEEG. In particular, we have demonstrated the first approach that
ERASE can potentially remove the confounding overlap between EMG and $\gamma$ signals, and preserve the expected brain signal features underlying motor behavior. The retained high-$\gamma$ activities demonstrated the expected increase during thumb flexion in contralateral hand motor cortex area (within the hemicraniectomy site). Moreover, the relative FD of the EEG high-$\gamma$ from the HAs after applying the new approach was demonstrated to be strongly correlated to the amplitude of thumb flexion force. Therefore, this approach may allow researchers to confidently use the resulting high-$\gamma$ signals for subsequent analysis or practical applications.

\section{Appendix}
The appendix includes the contents about the z-scored power of $\mu$ and high-$\gamma$ in different conditions (baseline, after ERASE and after conventional ICA) for Subject 2-6 (Figs. 1-5), electrodes with significant correlation in different conditions for Subjects 2-6 (Figs. 6-10), and the 2D image of the electrode locations (Fig. 22).





%

\newcommand{\newblock}{}
\bibliographystyle{unsrt}
\bibliography{reference}


\section{Supplementary}
\begin{figure}[!ht]
  \centering
  \includegraphics[width=\textwidth]{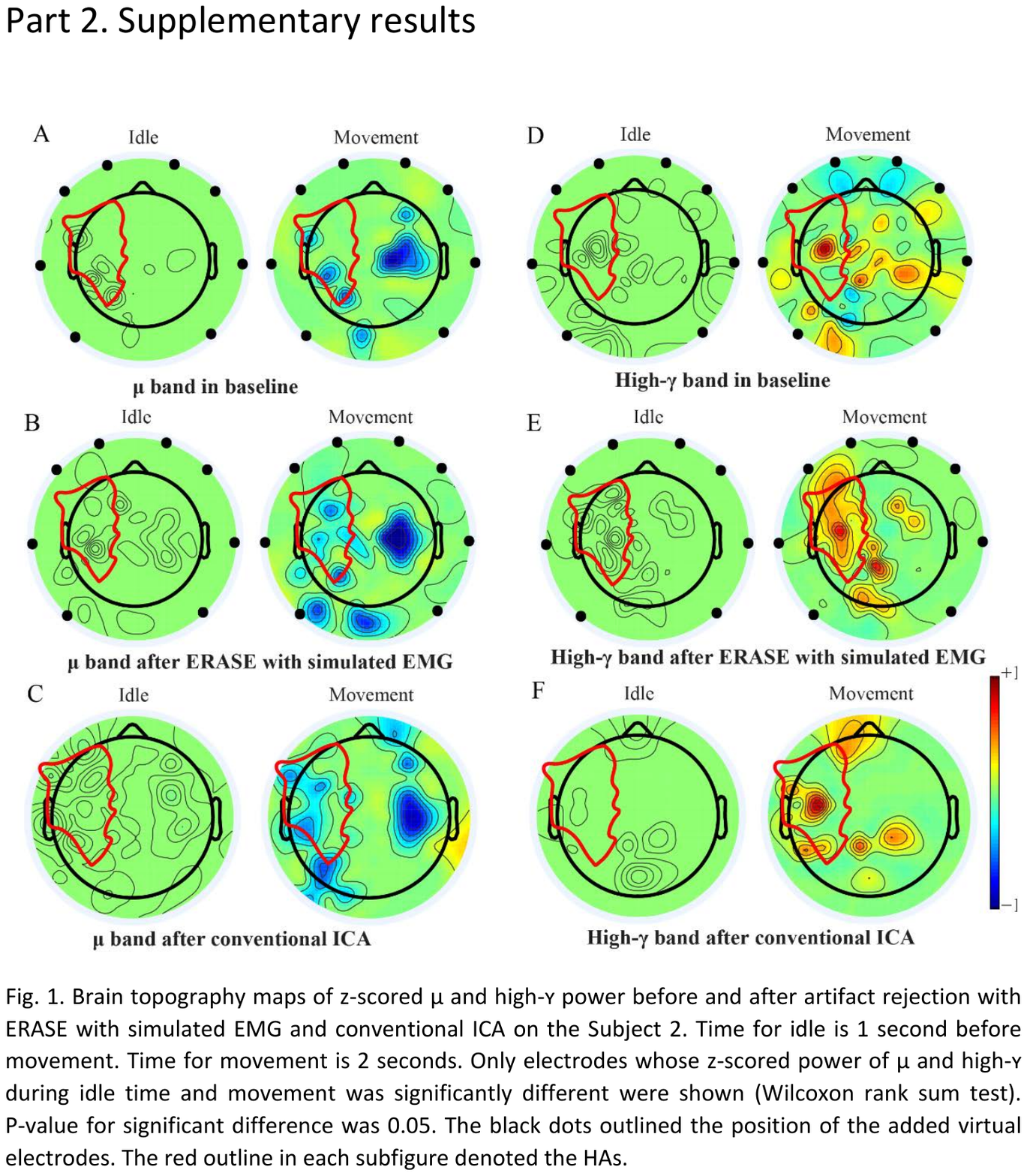}
\end{figure}

\begin{figure}[!ht]
  \centering
  \includegraphics[width=\textwidth]{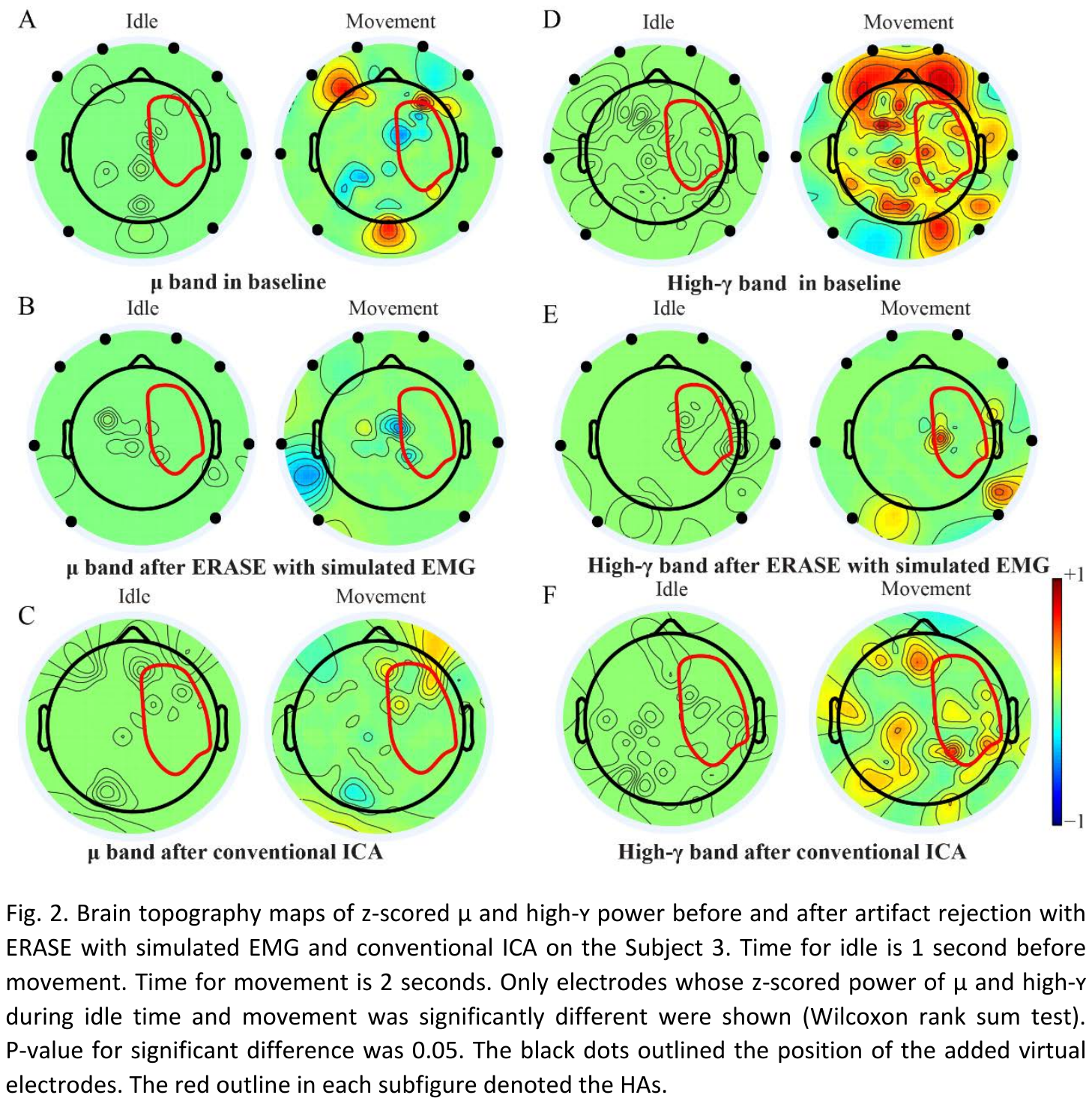}
\end{figure}

\begin{figure}[!ht]
  \centering
  \includegraphics[width=\textwidth]{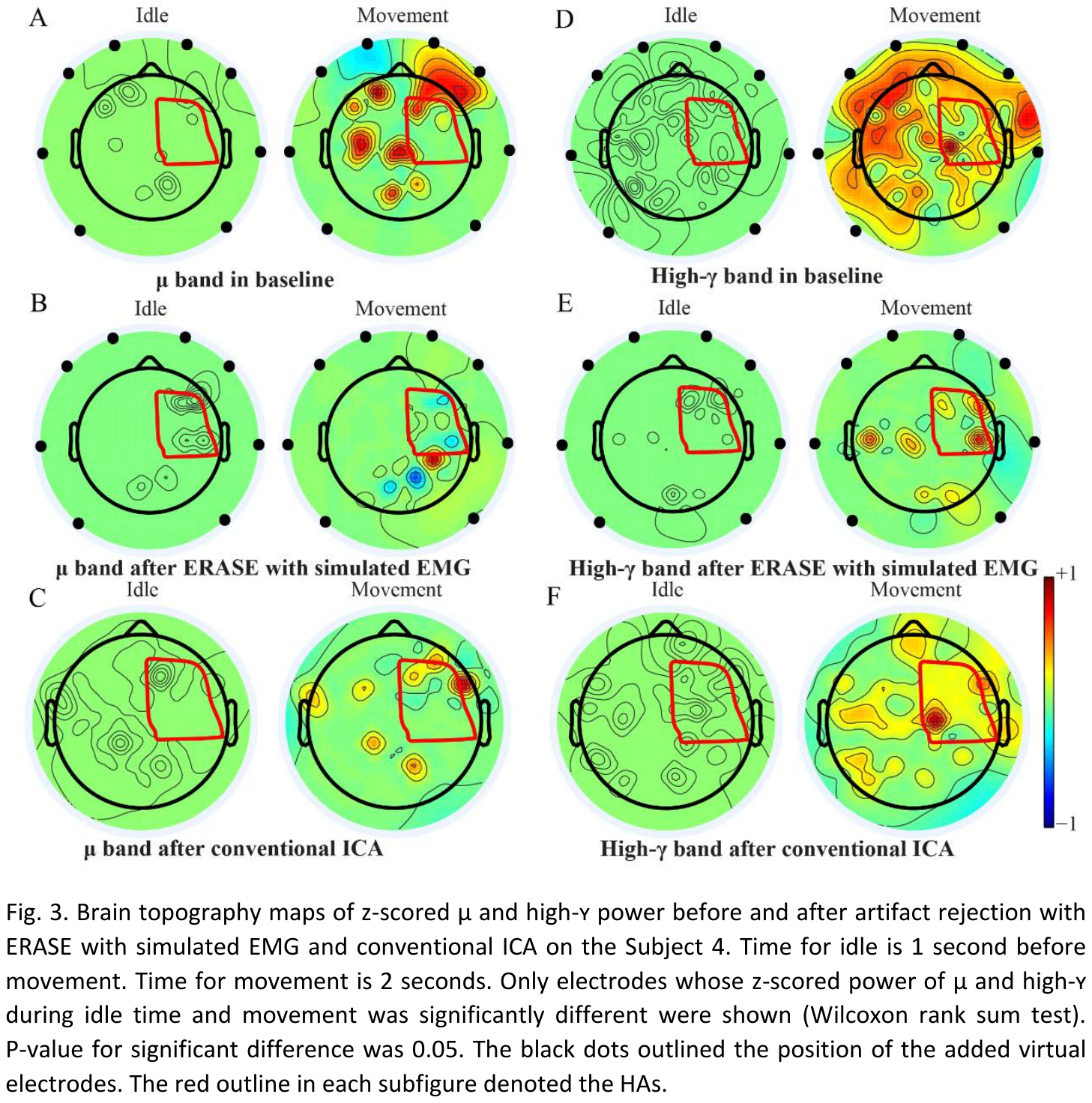}
\end{figure}

\begin{figure}[!ht]
  \centering
  \includegraphics[width=\textwidth]{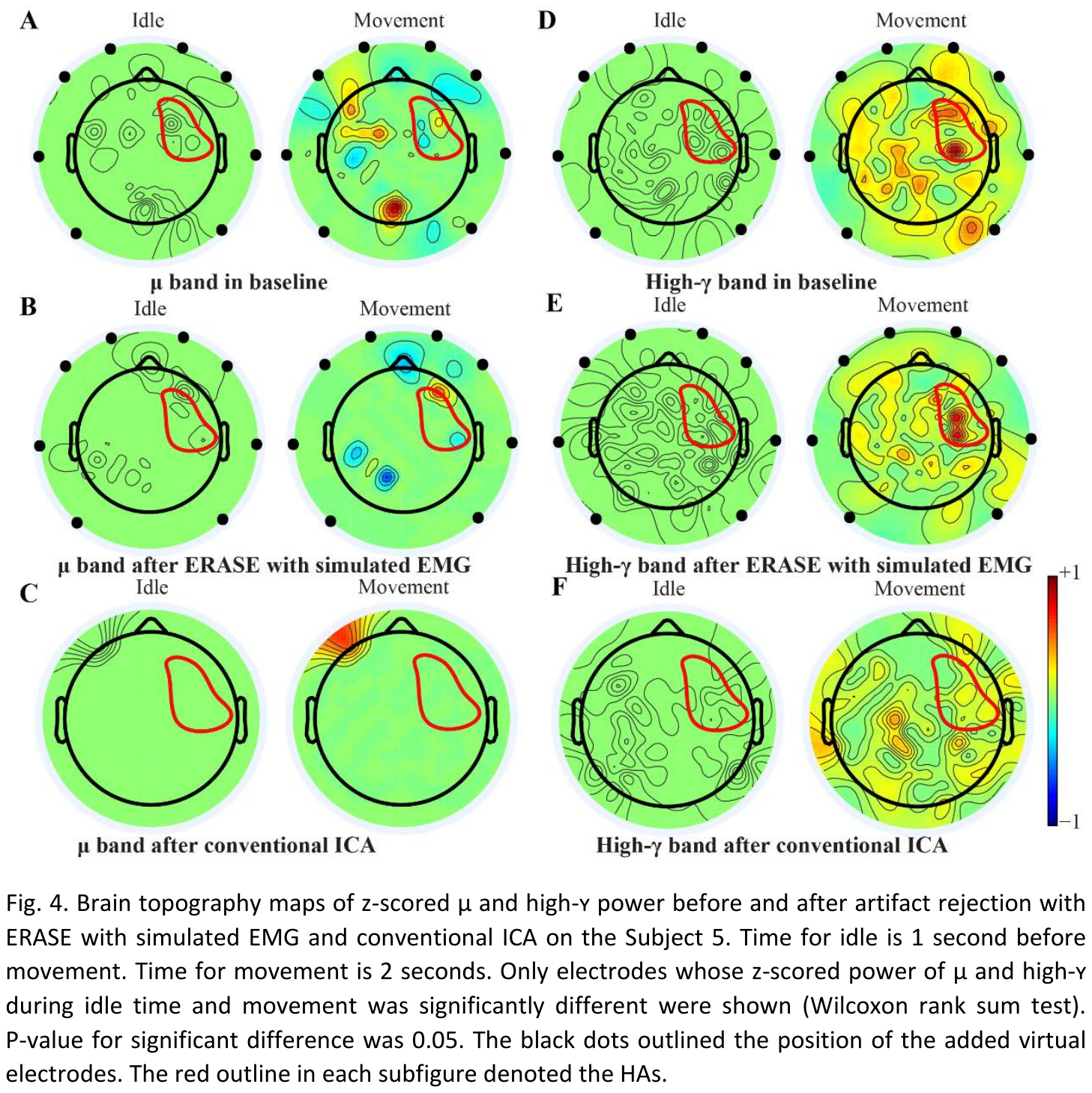}
\end{figure}

\begin{figure}[!ht]
  \centering
  \includegraphics[width=\textwidth]{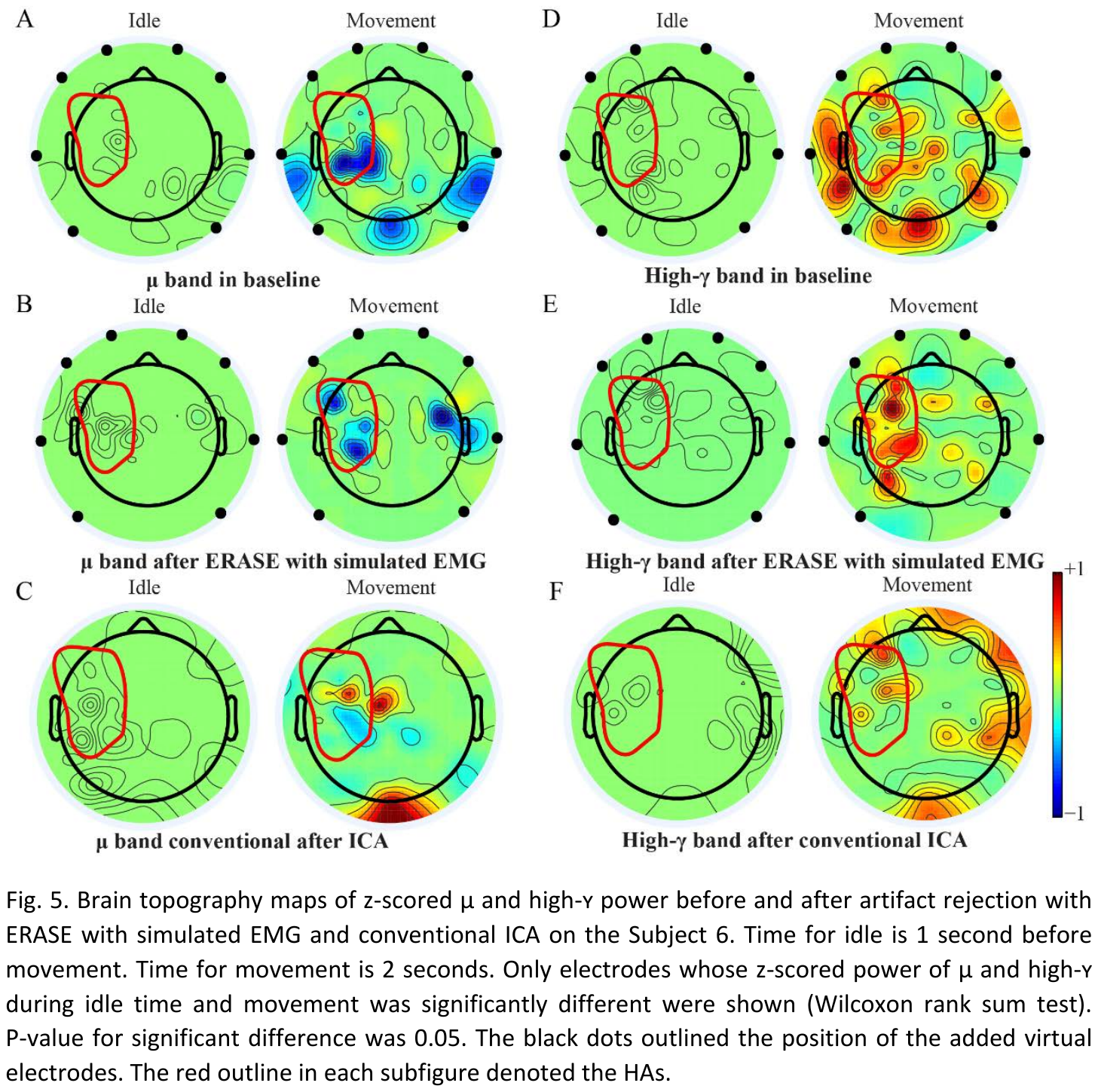}
\end{figure}

\begin{figure}[!ht]
  \centering
  \includegraphics[width=\textwidth]{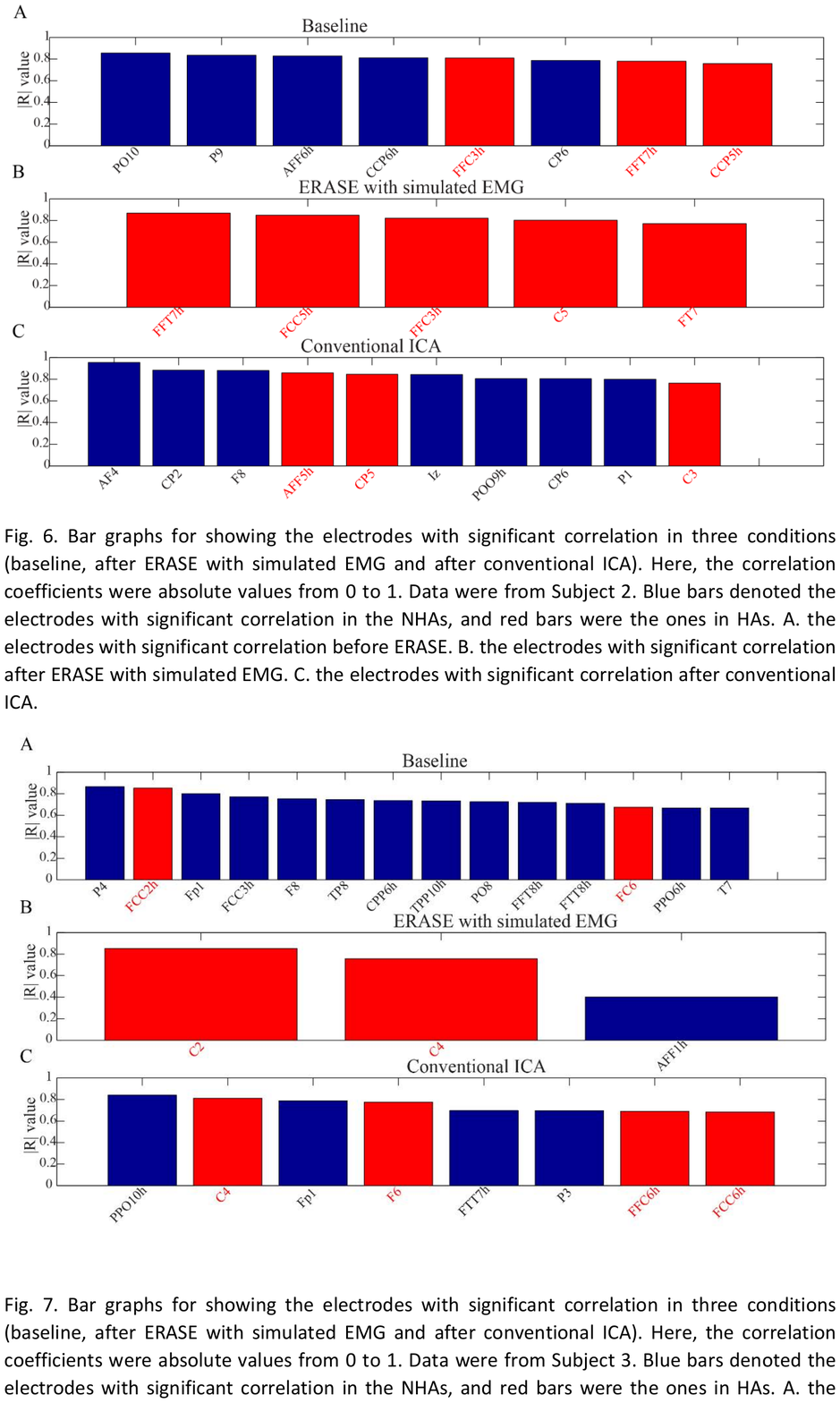}
\end{figure}

\begin{figure}[!ht]
  \centering
  \includegraphics[width=\textwidth]{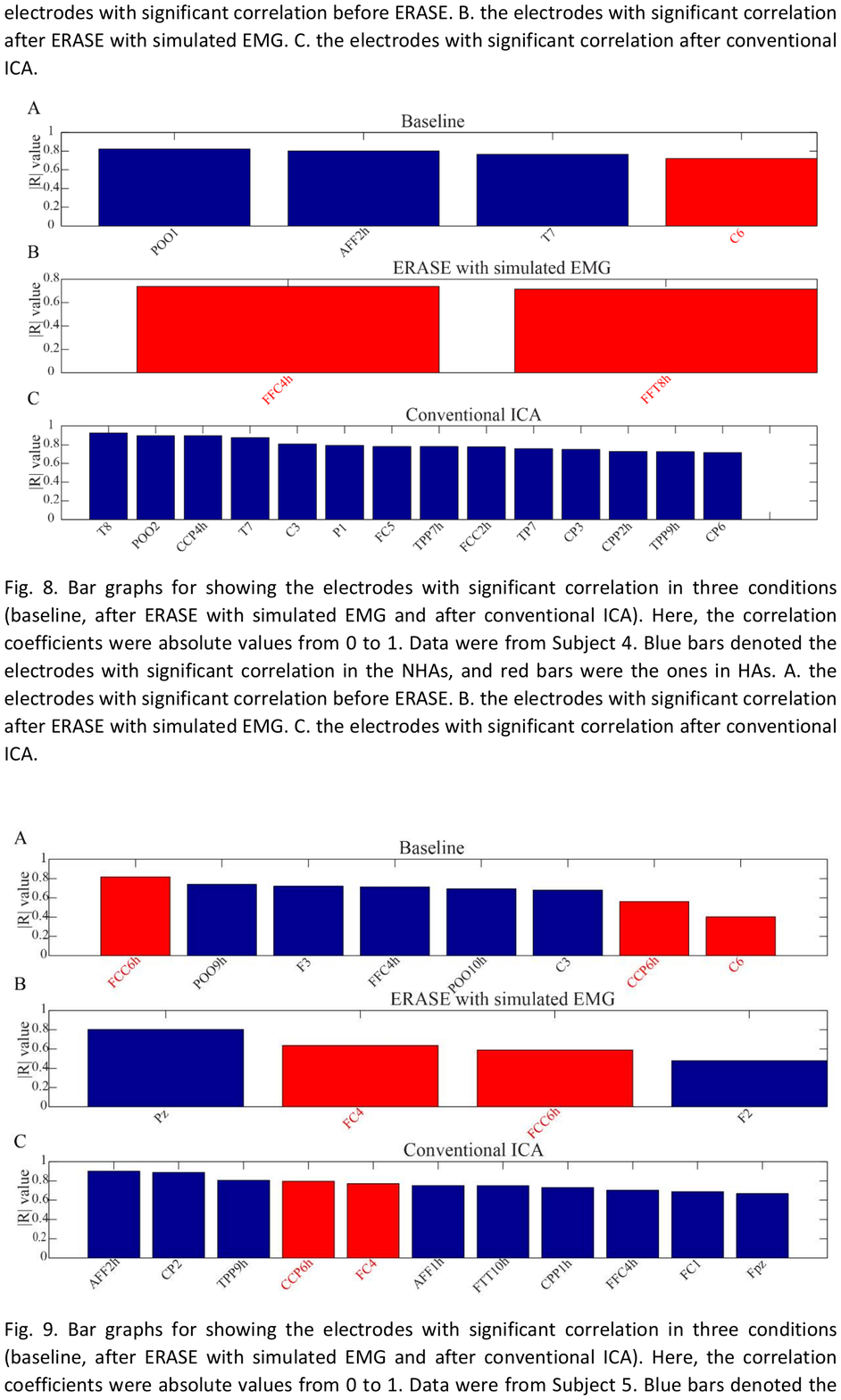}
\end{figure}

\begin{figure}[!ht]
  \centering
  \includegraphics[width=\textwidth]{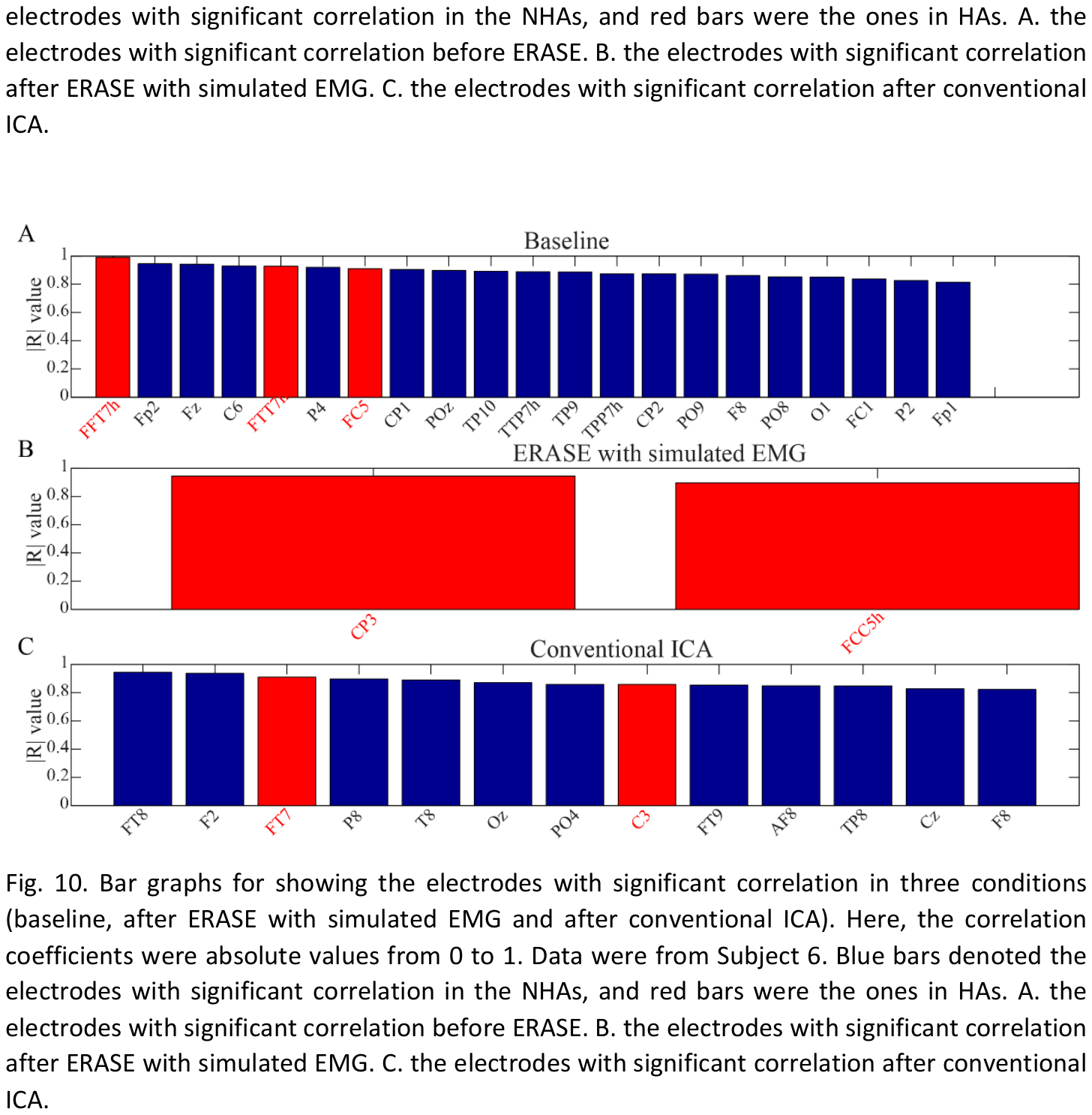}
\end{figure}

\begin{figure}[!ht]
  \centering
  \includegraphics[width=\textwidth]{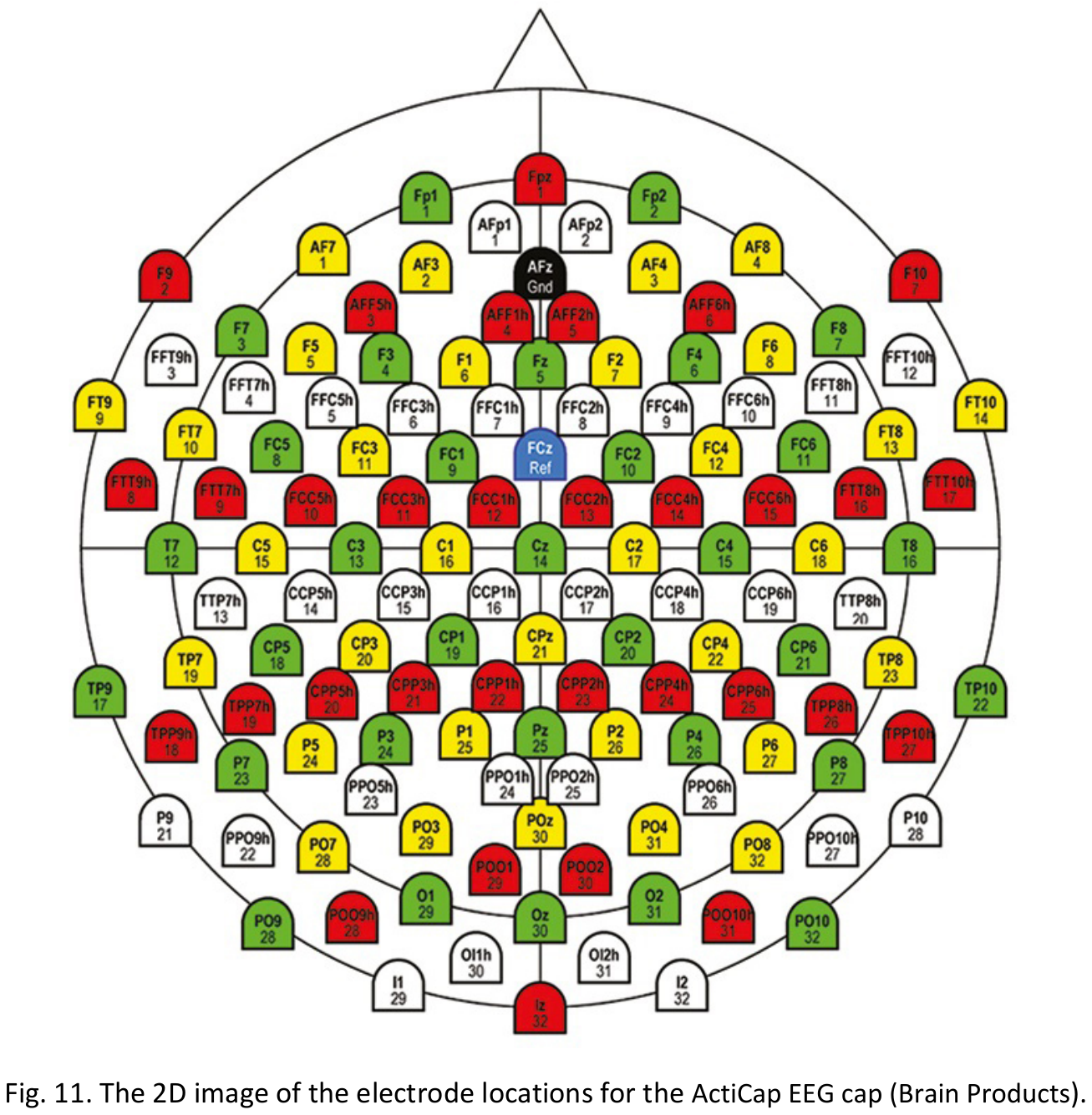}
\end{figure}
\end{document}